\documentstyle[editedbook,psfig,epsf]{mq}

\def\etal{{\em et al.} }

\def\msun{$M_{\odot}$ }

\def\cm2{cm$^2$ }
\def\se1{s$^{-1}$ }

\begin{opening}
\title{X-ray QPOs in Black-Hole Binary Systems}
\author{R. Remillard$^1$, M. Muno$^1$, J. McClintock$^2$ \& J. Orosz$^3$}
\institute{$^1$ Center for Space Research, M.I.T., 77 Massachusetts Ave.,
Cambridge, MA 02139\\
$^2$ Harvard-Smithsonian Center for Astrophysics, 60 Garden St., 
Cambridge, MA 02139\\
$^3$ Astronomical Institute, Utrecht University, Postbus 80000, 
3508 TA Utrecht, The Netherlands.}
\end{opening}

\runningtitle{Workshop Proceedings}
\runningauthor{Remillard \etal}

\begin{document}
\vspace{-0.5cm}

\begin{abstract}
{\small We briefly review the properties and physical consequences of
quasiperiodic oscillations (QPOs) seen on many occasions in the X-ray
emission from black-hole binary systems. High frequency QPOs ($\nu >$
40 Hz) continue to be scrutinized as effects of general relativity,
with new attention to the role of resonances in their
formation. Low-frequency QPOs (0.05 to 30 Hz) exhibit complicated
behavior, with occasions of high amplitude and particular correlations
with some X-ray spectral parameters. QPO mechanisms are a requirement
for any physical model seeking to explain either (1) the non-thermal
X-ray spectrum that is commonly seen and is usually stronger than the
accretion disk at times of highest luminosity, or (2) the hard X-ray
spectrum evident when there is a steady type of radio jet. }
\end{abstract}

\section{Introduction}

Most of the brightest celestial X-ray sources recorded in a given year
are transient outbursts from black hole binaries in the Galaxy. The
eruptions are understood as a consequence of a low rate of mass
accretion from the companion star \cite{tan96}. The material gradually
fills the outer regions of an accretion disk surrounding the black
hole. When the disk surface density reaches a critical value, matter
spirals into the inner disk where it reaches X-ray emitting
temperatures before falling into the black hole event horizon
\cite{mey01}.  Such X-ray novae are the parent population for
identifying black holes that are remnants of massive stars. The mass
of the compact object (typically 5--15 \msun) is deduced from radial
velocity studies of the companion star. In most cases such
measurements are only possible when the system has returned to
relative quiescence, and the companion can be seen against the glare
of the hot gases in the disk. In establishing the nature of the
compact object, the critical argument is whether the mass exceeds the
upper limit ($\sim 3.0$ \msun) for the mass of a neutron star.  There
are now 17 ``dynamical black hole'' binaries in the Milky Way or LMC:
14 were first noticed as bright X-ray novae \cite{oro02}, while the
other 3 cases are persistent X-ray sources with O/B type companions.

Observations with the {\it Rossi} X-ray Timing Explorer (RXTE) have
pioneered efforts to further study black holes and their occasional
relativistic jets via broadband X-ray observations during active
states of accretion. The X-ray timing and spectral properties convey
information about physical processes that occur near the black hole
event horizon, and one of the primary research goals is to obtain
constraints on the black hole mass and spin using predictions of
general relativity (GR) in the strong-field regime. There is also the
need to understand accretion physics for each of the four distinct
emission states that are displayed by so many accreting black holes
systems. In this paper we describe recent advances in these topics,
with particular attention to the diverse forms of quasiperiodic
oscillations (QPOs) \cite{vdk00} that have been detected during the
extensive monitoring programs for X-ray transients conducted with
RXTE.

\section{High Frequency Oscillations in Accreting Black Hole Systems}

\begin{figure}[htb]
\centering
\psfig{file=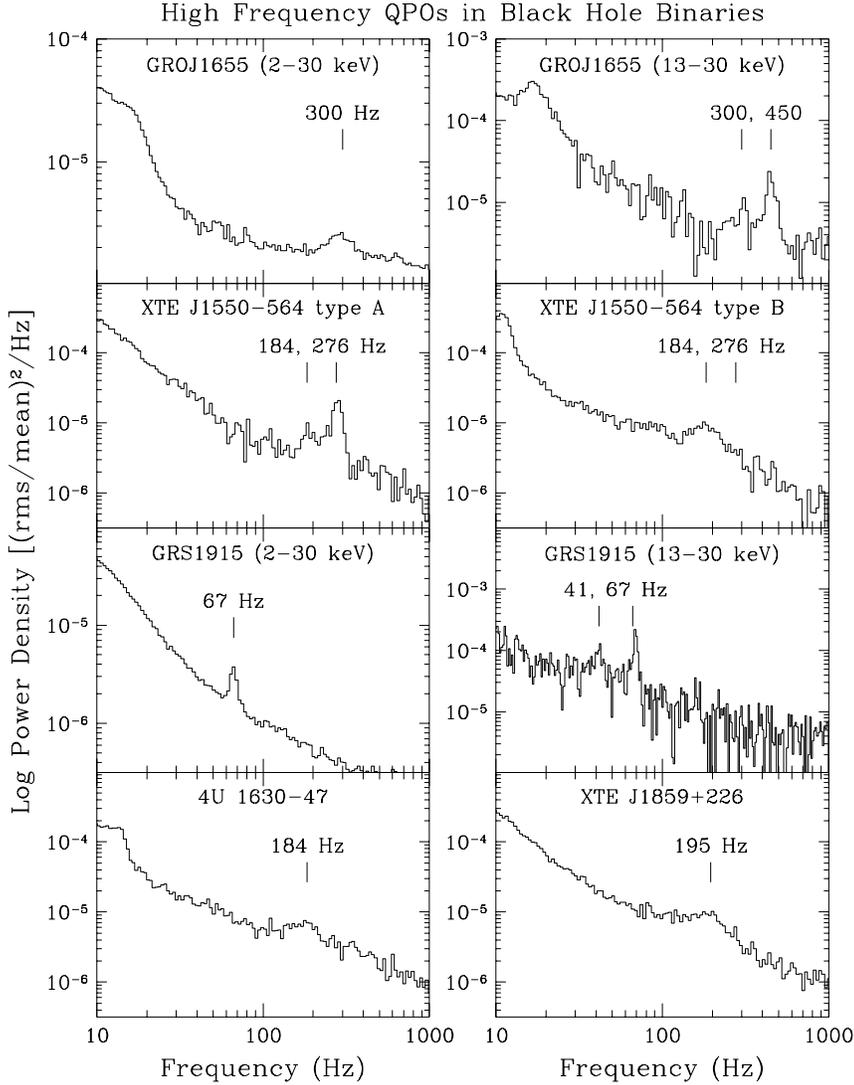,width=12cm}
\caption{HFQPOs in five black hole binary systems. The energy band is
6-30 keV unless otherwise indicated. These subtle oscillations are
only visible during a fraction of the observations for each source,
and details are given in Table 1.}
\label{fig:hfqpo}
\end{figure}

The topic of high-frequency QPOs (HFQPOs) in black hole binaries
(40-450 Hz) has evolved substantially in the last year.  Transient
HFQPOs are detected with RXTE\ in 5 sources (4 dynamical black holes
and 1 candidate). These subtle oscillations have rms amplitudes that
are typically only $\sim 1$ \% of the mean count rate.  The associated
power spectra are shown in Fig.~\ref{fig:hfqpo}, and their properties
are summarized in Table 1.  Three of these sources exhibit pairs of
HFQPOs: GRO~J1655--40 (300, 450 Hz; \cite{str01a}), XTE~J1550--564
(184, 276 Hz; \cite{mil01a,rem02a}), and GRS~1915+105 (40, 67;
\cite{str01b}). Furthermore, in Section 3 below we report the
detection of additional HFQPOs in GRS~1915+105 at 164 and 328 Hz.

The commensurate frequencies (3:2 ratio) in pairs of HFQPOs from \\
GRO~J1655--40 and XTE~J1550--564 \cite{rem02b} may suggest a cause in
some type of resonance phenomenon involving oscillations proscribed by
general relativity, as originally proposed by Abramowicz \& Kluzniak
\cite{abr01}. The observations cannot be interpreted as non-sinusoidal
structures in the wave pattern, in the sense of harmonics in a Fourier
series, because the individual detections (in a given energy band)
appear as a single peak in the power spectrum, and their frequencies
are recognized as $2 \nu$ or $3 \nu$ only when the ensemble of results
is examined.

\begin{table}[htbp]
\centering
\caption{Table 1: High Frequency QPOs in Black Hole Systems}
\begin{tabular}{|l r c c c r r|}
\hline
Source      & \# RXTE &  QPO Freq. &  Coherence    & Detections & Energy & Ampl. \\
Name        &  Obs.   & Hz ($\pm$) & ($\nu / FWHM$) & \# (3 $\sigma$)  & (keV) & rms \% \\
\hline
GRO~J1655--40  &  81  & 300 (23) & 4.4--11.6 &  7 &  2--30 & 0.5--1.0 \\
              &      & 450 (20) & 8.8--14.8 &  6 & 13--30 & 2.5--5.3 \\ 
              &      &          &           &    &        &        \\
XTE~J1550--564 & 364  & 184 (26) & 2.6--9.1  & 10 &  2--30 & 0.6--1.7 \\
              &      &          &           &    &        &        \\
              &      & 272 (20) & 4.7--14.3 & 14 &  6--30 & 3.6--6.9 \\ 
              &      &          &           &    &        &        \\
GRS~1915+105   &	448  &  67 (5)  & 3.2--22.0 & 29 &  2--30 & 0.4--1.8 \\
              &      &  41 (1)  &  5--20    &  5 & 13--30 & 1.8--3.2 \\
              &      & 164 (2)  &  5--7   & 2 sums[14] & 13--30 & 2.0--2.2 \\
              &      & 328 (4)  & 14--16  & 2 sums[14] & 13--30 & 1.0--1.3 \\ 
              &      &          &           &    &        &        \\
4U~1630--47   & 301  & 184 (5)  &  5--9   & sum[30] & 6--30 & 1.0--1.2 \\ 
              &      &          &           &    &        &        \\
XTE~J1859+226 & 135  & 193 (4)  &  3--5   & sum[48] & 6--30 & 1.7--2.0 \\
\hline
\end{tabular}
\label{table:qpos}
\end{table}

The resonance hypothesis \cite{abr01} has been discussed in terms of
accretion blobs following perturbed orbits in the inner accretion
disk.  Unlike Newtonian gravity, GR predicts independent oscillation
frequencies for each spatial coordinate for orbits around a rotating
compact object, as seen from the rest frame of a distant observer.  GR
coordinate frequencies and their differences (i.e. beat frequencies)
have been proposed to explain some of the X-ray QPOs from both neutron
stars and black holes \cite{ste99}.  At the radii in the accretion
disk where X-rays originate, the GR coordinate frequencies are
predicted to have varying, non-integral ratios. The commensurate
frequencies in pairs of HFQPOs, noted above, can therefore be
interpreted as a signature of resonance, e.g. between pairs of
coordinate frequencies in the inner disk.  Unlike the azimuthal and
polar coordinates frequencies, the radial coordinate frequency reaches
a maximum value and then falls to zero as the radius decreases toward
the location of the innermost stable orbit (see \cite{kat01},
\cite{mer01}, and references therein). This ensures the possibility of
commensurate coordinate frequencies somewhere in the inner disk. For
example, there is a wide range in the dimensionless spin parameter,
$a_*$, where one can find a particular radius that corresponds to a
2:1, 3:1, or 3:2 ratio in the orbital and radial coordinate
frequencies.  A resonance between the polar and radial coordinate
frequencies is also possible.  In the resonance concept, nonlinear
perturbations may grow at these radii, ultimately producing X-ray
oscillations that represent some combination of the individual
resonance frequencies, their sum, or their difference. However, there
remain serious uncertainties as to whether such structures could
overcome the severe damping forces and emit X-rays with sufficient
amplitude and coherence to produce the observed HFQPOs \cite{mar98}.

Models for ``diskoseismic'' oscillations adopt a more global view of
the inner disk as a GR resonance cavity \cite{kat80,wag99}. This
paradigm has certain attractions for explaining HFQPOs, but integral
harmonics are not predicted for the three types of diskoseismic modes
derived for adiabatic perturbations in a thin accretion disk.
Clearly, there is also a need to investigate the
possibility of resonances within the paradigm of diskoseismology.

As initially noted by Strohmayer \cite{str01a}, the HFQPO measurements
may be seen to suggest substantial black hole spin.
For the 3 sources with pairs QPOs, optical or IR studies have yielded
measurements of the black hole mass. In each case the fastest QPOs
exceed the maximum orbital rotation frequency ($\nu_{\phi}$) at the
innermost stable circular orbit around a Schwarzschild black hole
\cite{sha83} (i.e. dimensionless spin parameter, $a_* = 0$). If the
maximum Keplerian frequency is the highest frequency at which a QPO
can be seen, then the measurements require prograde spin, e.g. $a_* >
0.15$ for GRO~J1655-40.  On the other hand, even higher values of the
spin parameter are required if the QPOs represent either a resonance
in coordinate frequencies \cite{rem02a} ($a_* > 0.25$), or pairs of
diskoseismic modes that exhibit commensurate frequencies by chance
\cite{wag01} ($a_* > 0.9$).

Encouragement for linking HFQPOs and GR oscillations can be found in
the fact \cite{rem02b} that the HFQPOs in XTE~J1550--564 and
GRO~J1655--40 scale $\sim M^{-1}$. This result is generally consistent
with the known mechanisms related to disk oscillations in the
strong-field regime of general relativity, as long as the values of
the spin parameter ($a_*$) are similar for these two black
holes. These results illustrate both the quantitative value of HFQPO
detections as a means of probing the physical properties of black
holes, and also the need to continue efforts to independently measure
black hole masses via dynamical optical studies.  The spin
measurements would be of further value towards evaluating the role of
black hole rotation in the production of jets \cite {bla77}.

Finally, the investigation of HFQPOs and the the energy spectra for
both XTE~J1550--564 and GRO~J1655--40 show identical patterns of
behavior that demand explanation \cite{rem02b}. There is a systematic
increase in the strength of the hard X-ray power-law component
(i.e. the type with index $\sim 2.5$) as the observed HFQPO shifts
from $3 \nu$ to $2 \nu$. We also note that HFQPOs are generally not
detected either when the accretion disk dominates the spectrum
(i.e. power-law is absent), nor when there is radio emission and the
power-law is flatter (index $\sim 1.7$).  These patterns of spectral
behavior were used to select the data that yields a new HFQPO
detection for GRS~1915+105, as described in the next section.

\section{New HFQPOs with Harmonics in GRS~1915+105}

GRS~1915+105 is the most extraordinarily variable X-ray binary known
\cite{bel00}.  It is also the first member of the ``microquasars'', a
handful of Galactic X-ray sources that produce relativistic jets
\cite{mir99}, as seen in the ``superluminal'' separation of bipolar
knots in sequences of radio images.  The system is a confirmed black
hole binary that contains a $\sim 14$ \msun black hole \cite{gre01}. A
rich variety of correlations have been found between X-ray, radio, and
infrared outbursts \cite{mir99,eik98,fen99a,fen99b,dha00,cor00},
contributing substantially to the overall success in linking mass
ejections to conditions in the inner accretion disk.

There is an extensive archive of $\sim 800$ pointed observations of
GRS~1915+105 obtained with RXTE during the last six years.  About half
of these show substantial variability, while the others show roughly
steady flux with $\sigma / \mu < 0.18$, evaluated in 1 s bins.
Studies of the energy division between the thermal/disk and hard
power-law components suggest that most of the steady states are either
hard states with persistent radio emission or soft states in which
most of the flux originates in the disk \cite{mun99}.  The transient
HFQPO at 67 Hz \cite{mor97} tends to appear in the latter condition,
especially when the spectrum is soft and very bright \cite{rem00}.  On
the other hand, the steady states of GRS~1915+105 offer limited
exposures to conditions when {\bf both} the disk and power-law exhibit
high luminosity, which corresponds to the times when harmonic HFQPOs
appear in XTE~J1550-564 and GRO~J1655-40. We are therefore motivated
to investigate the variable states of GRS~1915+105, using a scheme
to sort the exposures into quadrants of the color-intensity diagram,
in order to accumulate significant exposure time when the source is
bright and moderately hard.

\begin{figure}[htb]
\centering \psfig{file=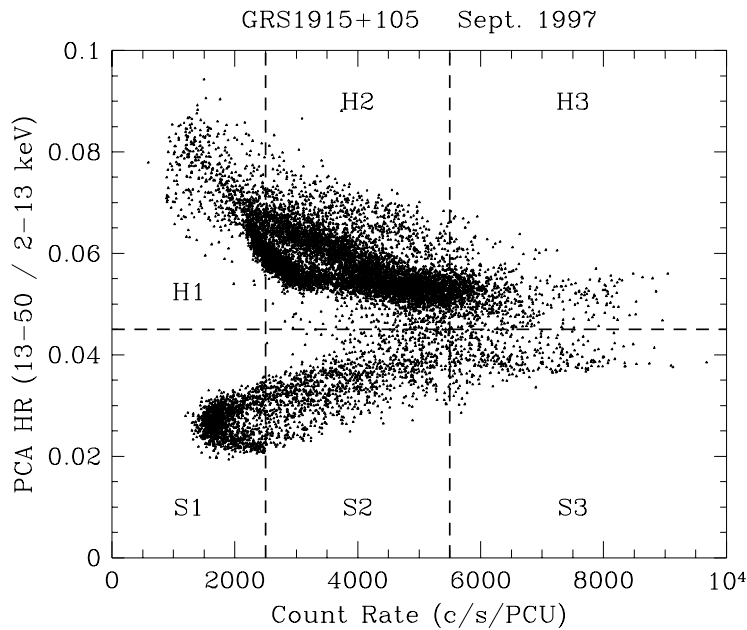,width=8cm}
\caption{Color-intensity diagram for GRS~1915+105 using RXTE PCA
observations on 10 days during 1997 September 5-29. The data from all
operating PCUs have been normalized and combined for each 16 s
exposure. The dashed lines illustrate the sorting boundaries used to
extract times when the source has different levels of the intensity
and the relative strength of the power-law component.}
\label{fig:1915ci}
\end{figure}

\begin{figure}[htb]
\centering \psfig{file=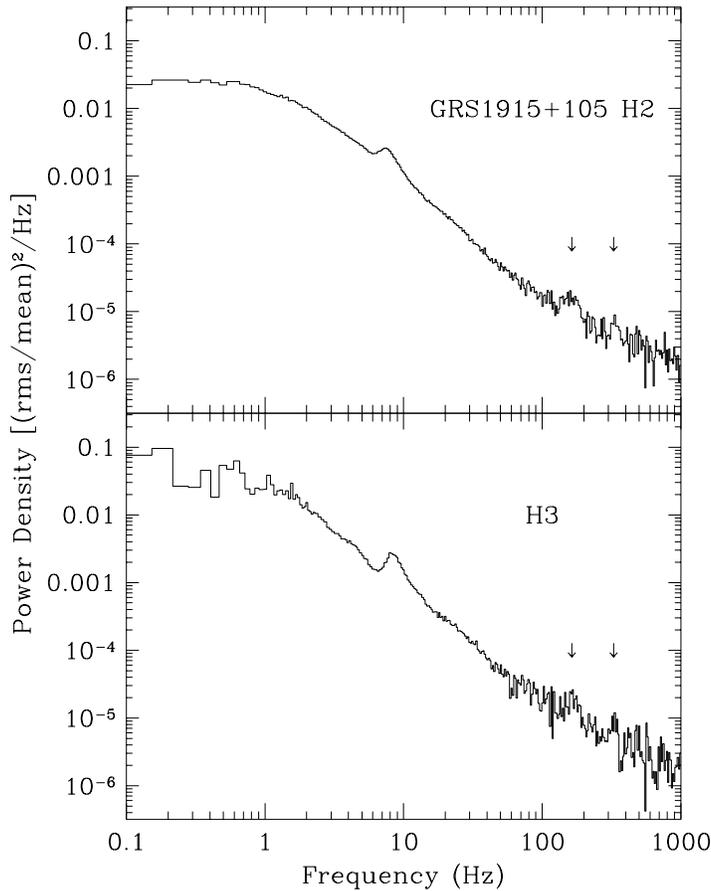,width=8cm}
\caption{Average power spectrum for groups H2 and H3 (see Fig. 2),
selecting only the PCA detections above 13 keV.  The arrows indicate
weak features at 164 Hz and the locations of the next harmonics
(i.e. 328 and 492 Hz). }
\label{fig:1915pds}
\end{figure}

For the present analysis, we select the 14 RXTE observations during
1997 September 5-29. These data offer long exposures, some rare
bright-hard-steady states, and types of wild variations that include
frequent excursions into bright-hard states.  The latter cases are
highlighted by light curves of the $\theta$ classification of Belloni
et al. \cite{bel00}.  The background-subtracted PCA spectra are
analyzed at a time resolutions of 16 s. The output of all operating
PCUs is combined for each time interval, after normalizing each PCU to
the characteristics of PCU \#2. The PCU normalization is defined using
six energy intervals and simply scaling the mean count rate per
interval to the results for PCU \#2, using the mean spectrum derived
for the subset of the observations in which both PCUs were
operating. Locations on the color-intensity diagram are computed using
a hardness ratio ($HR$) of normalized source counts at 13-30 keV
vs. 2-13 keV.  The resulting color-intensity diagram and the data
sorting scheme are shown in Fig~\ref{fig:1915ci}. The hard (``H'') and
soft (``S'') regions are separated by the line, $HR = 0.045$, and there
are three intensity levels (1,2,3), as shown in the Figure.

\begin{figure}[htb]
\centering \psfig{file=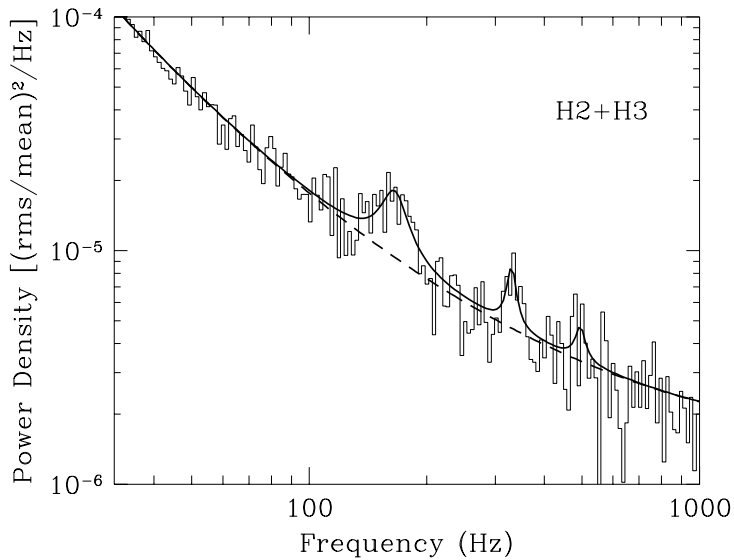,width=8cm}
\caption{Average power spectrum for groups H2 and H3 (13-30 keV)
with the best fit for QPO features and the power continuum. The QPO at 164.9 Hz
has a significance of 9 $\sigma$. The QPO at the first harmonic (329.8 Hz) is
detected at 3.7 $\sigma$, while the hint of a second harmonic (484.8 Hz)
is not statistically significant (1.5 $\sigma$). }
\label{fig:1915fit}
\end{figure}

The average power density spectra (PDS) are then computed for the data
in each of the six regions of the color-intensity diagram.  The search
for HFQPOs (see methods in \cite{rem02b}) produces detections
restricted to regions H2 and H3, using data confined to the energy
range of 13-30 keV. These PDS are shown in Fig~\ref{fig:1915pds}. The
H2 PDS shows a peak with a significance of 5.7 $\sigma$ at $163 \pm 5$
Hz with an rms amplitude of $1.56 \pm 0.14$ \% of the mean count rate
at 13-30 keV. Similarly, the H3 PDS yields a 4.4 $\sigma$ HFQPO at
$166 \pm 3$ Hz with amplitude $1.73 \pm 0.20$ \%.

We then combine the H2 and H3 PDS (weighted mean values per bin) and
fit the results for an HFQPO with harmonics. The data and best-fit
model are shown in Fig~\ref{fig:1915fit}, and the HFQPO parameters are
included in Table 1. With the first harmonic included in the model,
the HFQPO at $165 \pm 3$ Hz has a significance above 9 $\sigma$. The
second peak at 330 Hz has a significance of 3.7 $\sigma$, while there
is a hint (not significant) of the next harmonic at 495 Hz.

In conclusion, the spectral characteristics of HFQPOs in XTE~J1550-564
and GRO~J1655-40 have provided guidance for finding new HFPQOs at 165
and 330 Hz in GRS~1915+105. The 165 Hz QPO in GRS~1915+105 then
appears closely related to the QPO systems with $\nu_0 = 150 $ Hz and
92 Hz in GRO~J1655-40 and XTE J1550-564, respectively. Given the
optical mass estimates for all three black holes, the results suggest
faster spin in GRS~1915+105, compared to the other two sources, if the
QPOs originate from a common inner disk oscillation related to general
relativity. On the other hand, the QPOs at 41 and 67 Hz in
GRS~1915+105 may arise from a different physical mechanism.

\section{Low Frequency QPOs and Problems in Accretion Physics}

The X-ray PDS of many black hole transients display low frequency QPOs
(LFQPOs), roughly in the range of 0.05 to 30 Hz. The oscillations can be
particularly strong, with peak to trough ratios as high as 1.5
\cite{sob00b}. LFQPOs have been traced out to energy bands as high as 
60-124 keV \cite{tom01}. The measurement properties of LFQPOs are summarized
in Table 2. Some authors pursue separate interpretations for HFQPOs
and the more common LFQPOs because they seem to evolve separately and
may sometimes coexist \cite{rem02a}. However, others argue that there
are unification schemes involving both neutron stars and black holes
in which particular mechanisms may operate across a wide range in
frequency \cite{psa99}.

Investigations of LFQPO properties versus spectral parameters of the
disk and power law components generally show that the LFQPO frequency
is most often correlated with the disk flux \cite{tru99,sob00a}.
However, the rms amplitude generally increases between 2 and 20
keV, and the amplitude spectrum therefore resembles the power-law
component rather than the accretion disk \cite{tom01}. This
dualism in correlations encourages LFQPO investigations 
motivated to deduce the origin of the power-law spectrum.  One major
complication here is that broad-band spectral studies tend to
emphasize the importance of distinguishing two forms of X-ray power
law spectra \cite{gro98}. At highest luminosity (the ``very high''
state), the spectrum is dominated by a power law with a photon index,
$\Gamma \sim 2.5$, which may extend to 500 keV or higher \cite{tom99}.
Alternatively, in the ``low/hard'' state, the power-law component
shows a flatter spectrum ($\Gamma \sim 1.7$) that is cut off above
$\sim 100$ keV, and this state is almost always associated with a
steady radio jet\cite{fen01}.  On the other hand, the PDS and QPO
properties of these two states may appear rather similar
\cite{mun01,sob00b}.

\begin{table}[htbp]
\centering
\caption{Table 2: Properties of Low Frequency QPOs in Black Hole Systems}
\begin{tabular}{|l c r|}
\hline
Property         &     Value      &   Comments \\
\hline
Frequency range  &  0.05 -- 30 Hz &  most 1--7 Hz \\
rms amplitude    &  1 -- 20 \%    &  2-30 keV; higher at 6--30 keV \\
$Q=\nu/\Delta\nu_{FWHM}$ &  3--30   &  typical $\sim 8.5$  \\
Phase Lag        &  -0.1 -- 0.2   &  2-6 keV vs. 13-30 keV \\
Harmonics        &  sometimes 2 and 0.5 $\nu_0$  &  e.g. XTE~J1550 and GRS~1915 \\
  &  &  \\
In Low-Hard State  &    often     &  $\Gamma_{PL} \sim 1.7$ + radio flux \\
In High-Soft State &      no      &  thermal spectrum, weak power law \\
In Very High State &     yes      &  defining item; $\Gamma_{PL} \sim 2.5$ \\
  &  &  \\
Min. Turn-On Timescale   & seconds &  GRS~1915+105 light curves \\
Max. Longevity Timescale & months  &  GRS~1915+105 low-hard state \\
Freq. Correlation & with Disk Flux     & all other params worse \\
Ampl. Correlation & with Photon Energy & like power-law spectrum, not disk \\
\hline
\end{tabular}
\label{table:lfqpos}
\end{table}

For the strongest QPOs in GRS~1915+105, efforts were made to track the
individual oscillations to determine the origin of the frequency
drifts and to measure the average ``QPO-folded'' oscillation profile
\cite{mor97}. The results show a random walk in QPO phase, with only
minor departures from a sinusoidal waveform. The ramifications of this
study for QPO models remain uncertain. There are now a large number of
proposed LFQPO mechanisms in the literature, and a full review is
beyond the scope of this paper.  The models include global disk
oscillations \cite{tit00} as well as oscillations in the radial
position of complex accretion structures, such as shock fronts
\cite{cha00} or a transition layer between the disk and a hotter
Comptonizing region \cite{nob00}. The construction of complex
accretion models is driven not only by the need to account for the
power-law spectrum, but also by the relatively slow frequencies of
LFQPOs.  Given the capacity of LFQPOs for high amplitudes and
long-term stability (see Table 2), it would be attractive to associate
them with Keplerian frequencies in the disk; however, the relevant
disk radii are well beyond the expected limits of the X-ray emitting
region. An alternative model known as the ``accretion-ejection
instability'' invokes spiral waves in a magnetized disk \cite{tag99},
with a transfer of energy out to the radius of corotation with the
spiral wave. This model uses magnetic field effects to account for
QPOs, the power-law spectrum, and perhaps relativistic outflows as
well. Further details are given in other papers of this workshop.

\begin{figure}[htb]
\centering \psfig{file=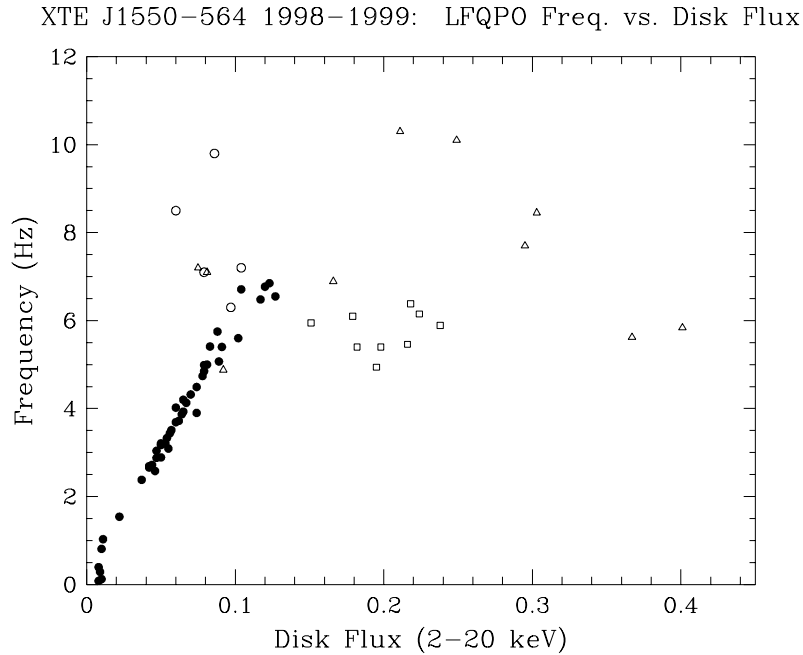,width=9cm}
\caption{Correlation between LFQPO frequency and the apparent disk
flux for XTE~J1550-564 during its 1998-1999 outburst. The plotting
symbols distinguish the QPO type: Type A (broad QPOs with phase lags
in soft X-rays) -- open triangles, Type B (narrow QPOs with hard lags)
-- open squares, Type C (small lags and strong harmonics) -- filled
circles, and anomalous QPOs -- `x'. The correlation between these
quantities is only demonstrated for the C type LFQPOS. }
\label{fig:1550lfqpos}
\end{figure}

Recent efforts to understand LFQPOs have given attention to the phase
lags associated with these oscillations and their harmonics. The
analysis technique uses Fourier cross spectra to measure both the
phase lags and the coherence parameter (versus frequency) between
different X-ray energy bands, e.g. 2-6 vs. 13-30 keV.  Unexpectedly,
both positive and negative phase lags were found
\cite{wij99a,cui00,rei00,mun01}, and suggestions were made to classify
LFQPOs by phase lag properties \cite{wij99a}. While the expansion of
LFQPO subtypes may not be a welcome complication, it has been shown
that recognition of the phase lag properties allows us to better
understand the relationship between certain types of QPOs and the
conditions in the disk \cite{rem02a}.  This is illustrated in
Fig~\ref{fig:1550lfqpos}.  Furthermore, data selections by LFQPO
subtypes provided guidance that helped to establish the harmonic
nature of HFQPOs in XTE~J1550-564 and the relationship between LFQPOs
and HFPQOs in that source \cite{rem02b}.

In closing, it must be noted that the relevance of X-ray timing
studies requires discussions of broad power peaks in the PDS
\cite{now00,bel02}, which are features wider than the limit ($Q \sim
2$) used to define QPOs.  Broad power peaks are involved in important
studies such as the means to distinguish accreting black holes from
neutron stars \cite{sun00}. The evolution of broad power peaks has
also been linked to major behavioral changes in a black hole binary,
as shown in disappearance of a broad power feature just at the time
when Cyg X-1 recently lost its ability to remain in the low-hard state
and maintain a steady jet \cite{pot02}.

\end{document}